\documentclass[aps,twocolumn,groupedaddress]{revtex4-1}

\usepackage{graphicx}% Include figure files
\usepackage{dcolumn}% Align table columns on decimal point
\usepackage{bm}% bold math
\usepackage{amsfonts}
\usepackage{amsmath}
\usepackage{amssymb}
\usepackage{xcolor}
\usepackage{enumerate}

\newcommand{\can}{Ca$_2$N}
\newcommand{\ban}{Ba$_2$N}
\newcommand{\bap}{Ba$_2$P}
\newcommand{\srn}{Sr$_2$N}
\newcommand{\srp}{Sr$_2$P}
\newcommand{\yc}{Y$_2$C}
\newcommand{\baas}{Ba$_2$As}
\newcommand{\ab}{$A_2B$}
\newcommand{\abab}{$(A_2B)_2$}
\newcommand{\oab}{O/$A_2B$}
\newcommand{\oababo}{O/$(A_2B)_2$/O}
\newcommand{\obanbanot}{(BaO(BaN)$_2$BaO)$^{\rm t}$}
\newcommand{\ocancanot}{(CaO(CaN)$_2$CaO)$^{\rm t}$}
\newcommand{\osrnsrnot}{(SrO(SrN)$_2$SrO)$^{\rm t}$}
\newcommand{\oabt}{($A$O$AB$)$^{\rm t}$}
\newcommand{\oabh}{O/$(A_2B)^{\rm h}$}
\newcommand{\oababot}{$(A\text{O}(AB)_2A\text{O})^{\rm t}$}
\newcommand{\oaNt}{($A$O$A${N})$^{\rm t}$}
\newcommand{\oaN}{O/$A_2${N}}
\newcommand{\ocant}{(CaOCaN)$^{\rm t}$}
\newcommand{\ocan}{O/Ca$_2$N}
\newcommand{\ocancano}{O/(Ca$_2$N)$_2$/O}
\newcommand{\osrnt}{(SrOSrN)$^{\rm t}$}
\newcommand{\obant}{(BaOBaN)$^{\rm t}$}
\newcommand{\oaNaNot}{$(A$O($A$N)$_2$$A$O)$^{\rm t}$}

\newcommand{\oaNaNo}{O/($A_2\text{N})_2$/O}

\setcitestyle{super}
\begin{document}

\title{Oxidation of  2D electrides: structural transition and \\ the  formation 
of  half-metallic channels protected by oxide layers}

\author{Pedro H. Souza$^1$}
\email{psouza8628@gmail.com}

\author{Danilo Kuritza$^2$}
\email{danilo.kuritza@gmail.com}

\author{Jos\'e E. Padilha$^2$}
\email{jose.padilha@ufpr.br}

\author{Roberto H. Miwa$^1$}
\email{hiroki@ufu.br}

\affiliation{$^1$Instituto de F\'isica, Universidade Federal de Uberl\^andia,
        C.P. 593, 38400-902, Uberl\^andia, MG, Brazil}
        
\affiliation{$^2$Campus Avan\c cado Jandaia do Sul, Universidade Federal do 
Paran\'a, \\ 86900-000, Jandaia do Sul, PR, Brazil.}

\date{\today}
  
\begin{abstract}
\vspace{3mm}
\begin{center}
 {\bf ABSTRACT}
\end{center}

Based on first-principles calculations we performed a systematic study  of the 
energetic  stability, structural characterization, and  electronic properties of 
the fully oxidized \ab\, electrenes, with the following combinations, (i) $A$\,=\,Ca, Sr, and Ba for $B$\,=\,N; (ii) $A$\,=\,Sr and Ba for $B$\,=\,P; and  \yc\, and \baas. We have considered one side oxidation of  single layer 
electrenes (\oab),  and two side oxidation of bilayer electrenes (\oababo). We 
show  that the hexagonal lattice of the pristine host is no longer  the ground 
state structure in the (fully) oxidized   systems. Our total energy results 
reveal an exothermic structural transition from hexagonal to tetragonal 
(h\,$\rightarrow$\,t) geometry, resulting  in  layered tetragonal structures 
[\oabt\, and \oababot]. Phonon spectra calculations and molecular dynamic 
simulations show that the \oab\, and \oababo\, systems, with  $A$\,=\,Ba, Ca, 
Sr, and $B$=\,N, become dynamically and structurally stable upon such a 
h\,$\rightarrow$\,t transition. Further structural characterizations were 
performed based on simulations of the near edge X-ray absorption spectroscopy at 
the nitrogen K-edge. Finally, the  electronic band structure and 
transport calculations reveal the formation of half-metallic bands  spreading 
out through the $A$N layers, which in turn are shielded by oxide $A$O sheets.  
These findings indicate that \oaNt\, and \oaNaNot\,   are  
quite interesting  platforms for application in spintronics; since the 
half-metallic channels  along the $A$N or $(A\text{N})_2$ layers (core) are 
protected  against the environment conditions by the oxidized $A\text{O}$ sheets 
(cover shells). 
\end{abstract}

\maketitle

\section{Introduction}

Research works  in  two-dimensional (2D)  materials  with different 
functionalities have been boosted in the last few year. Since the successful 
synthesis of graphene,\cite{novoselovScience2004} 2D materials have been 
considered as a new paradigm not only  in fundamental studies, like  the search 
of  topological phases,\cite{kanePRL2003,weeksPRX2011,acostaPRB2014} and 
tuneable  magnetic structures in 2D 
systems,\cite{songNatMat2019,liNatMat2019,morell2DMat2019,dominikePRMat2021} but 
also  addressing technological applications. For instance,  the development of 
new materials for (nano)electronic and spintronic devices like single layer  
field effect transistors,\cite{radisavljevicNatNanotech2011} and half-metals 
based on  transition metal dichalcogenides  and  
dihalides.\cite{tongAdvMat2017,kulishJMatChemC2017,gokougluMatResExpress2017, 
fengJMatChemC2018, wangNanoscaleHoriz2018}

Electrides are ionic crystals characterized by the presence of electrons 
not bonded to a particular nucleus. These electrons act as ions (with no 
nucleus) embedded within the crystal lattice, anionic 
electrons.\cite{druffelJMatChemC2017,liuJPhysChemC2020} Further experimental 
works revealed that these anionic electrons form nearly free electron gas (NFEG) 
confined between the stacked layers of the 
electride.\cite{leeNature2013,ohJACS2016,druffelJACS2016} It is worth noting 
that  there are other native inorganic electrides with the same lattice 
structure of \can, like \srn\cite{breseJSolStateChem1990} and 
\yc;\cite{zhangChemMat2014} meanwhile   other ones that also share the same lattice structure of \can\, have been predicted  
throughout  high-throughput computational 
simulations, for instance, \ban, \srp, \bap, \yc,  and \baas.\cite{tadaInorgChem2014,inoshitaPRX2014,zhangPRX2017computer,
zhouChemMat2019discovery}

By taking advantage of the layered structure of their 3D parents,  combined with a   suitable balance between strong (weak) intralayer (interlayer) binding interactions,\cite{mounetNatNanotech2018}  two dimensional materials can be obtained through exfoliation processes. Indeed, in a seminal work, Lee {\it et al.}\cite{leeNature2013} revealed the exfoliable nature,  and the two dimensional electronic confinement in \can\,  electrides. Further theoretical studies, based on first-principles calculations,  confirmed the exfoliable nature of \can, its electronic properties, and energetic/structural stability of the \can\, electrenes.\cite{zhaoJACS2014,druffelJACS2016,zengPRB2018}

Currently,  few layer systems of electrides (electrenes) have attracted research works  in fundamental   issues, like the search of topological phases throughout the  design of  kagome lattices on the electrene surface,\cite{zhangPRB2019kagome} as well as to  the development of electronic devices with hight carrier density and electronic mobility. Further, theoretical  studies have addressed the functionalization of \can\, electrenes mediated by atomic adsorption.\cite{liuMaterResExpress2018} Functionalization is a quite promising route in order to tailor the electronic and magnetic properties of 2D systems. For instance, the rise of  ferromagnetic (FM) phases upon full hydrogenation,\cite{qiuJPhysChemC2019} and oxidation of \can\, and \srn\, electrenes.\cite{wuJMMM2020,souzaJPhysChemC2020} On the other hand, it is important to stress that  the presence of anionic electrons makes the electrene surface very reactive,  which may lead to significant changes on the electronic and structural properties of the functionalized  systems, giving rise to a new set of physical properties to be exploited.

In this work, by means of first-principles calculations, we perform a systematic 
investigation of the energetic  stability, structural, and the 
electronic/magnetic properties of the  oxidized  \ab\, electrenes, with the following combinations, (i) $A$\,=\,Ca, Sr, and Ba for $B$\,=\,N; (ii) $A$\,=\,Sr and Ba for $B$\,=\,P; and  \yc\, and \baas. We have considered the fully 
oxidation of one surface side of single layer electrene (\oab), and two surface 
sides  in bilayer electrenes (\oababo). Our total energy results revealed a 
barrierless hexagonal (h) to tetragonal (t) structural transition, giving  rise 
to  layered tetragonal systems, namely \oabt\, and \oababot. The dynamical and 
structural stabilities were examined through a combination of phonon spectra 
calculations, and  molecular dynamic simulations. Based on the simulations of 
the (N K-edge) X-ray absorption  near edge spectroscopy (XANES), and
the projection of the electronic orbitals, we present  a detailed structural 
analysis of the oxidized systems, and fingerprints of their tetragonal 
geometries. Finally, focusing on the electronic/magnetic properties, we find 
that the oxidized \oabt, and \oababot\, systems are characterized by (i) an 
energetic preference for the FM phase, and (ii) the emergence of half-metallic 
channels along $A$N layers shielded  by oxidized $A$O shells (with $A$\,=\,Ca, 
Sr, and Ba).

\section{Computational details}

The calculations were performed by using  the density functional theory (DFT),\cite{kohn} as implemented in the computational codes Quantum-Espresso (QE)\,\cite{espresso} and Vienna Ab initio Simulation Package (VASP).\cite{vasp1,vasp2} We have considered the generalized gradient approximation of Perdew-Burke-Ernzerhof (GGA-PBE)\,\cite{PBE} for the exchange-correlation functional,  and the electron-ion interactions were described using norm-conserving pseudopotentials (PS),\cite{hamann2013optimized} and projected augmented wave (PAW)\cite{paw} in  the QE and VASP codes, respectively. The single layer and bilayer \ab\, electrenes were simulated using slab structures within the supercell approach, with a vacuum region of 18 and 22\,\AA, respectively, and surface periodicities of (1$\times$1), and ($\sqrt 2$\,$\times$\,$\sqrt 2$) for  hexagonal and tetragonal structures. Dipol  corrections have been included in order to suppress artificial electric fields (in asymmetric systems) across the slab.\cite{beng-99} The final atomic geometries, and  total energies were obtained  using the QE code, where the Kohn-Sham\,\cite{KS} orbitals, and the self-consistent total charge densities  were expanded in plane wave basis sets with energy cutoffs of 70 and 353\,Ry; the  Brillouin zone sampling was performed by using a  8$\times$8$\times$1 k-point mesh.\cite{mp,vasp-check} The atomic positions were relaxed until the residual forces were converged to within  5\,meV/\AA, and the structural relaxation (variable-cell) was performed within  a pressure convergence of 0.05\,Kbar. The long-range van der Waals (vdW) interactions were taken into account using the self-consistent vdW-DF approach.\cite{dionPRL2004,perezPRL2009,klimevsPRB2011}  

Further structural characterizations were performed through calculations of the 
X-ray absorption spectra combining the QE results  and 
Xspectra\cite{xas1,xas2,xas3} simulations. We have considered the  K-edge 
spectra of nitrogen atoms by using the Gauge-Including Projector Augmented-Wave 
(GIPAW)\,\cite{gipaw} method to calculate the dipolar cross section, 
$$
\sigma(\omega) \propto \sum_{n}|\langle\psi_n |{\bf\hat{\varepsilon}\cdot 
r}|\psi_{1s}\rangle|^2\delta(\epsilon_n - \epsilon_{1s} - \hbar\omega),
$$
within the dipole approximation; $\psi_n$/$\epsilon_n$ and 
$\psi_{1s}$/$\epsilon_{1s}$ are the  final $n$ 
and initial $1s$ (single particle) orbitals/energies in the presence of 
core-hole. The absorbing atom is 
described with a pseudopotential with a full core-hole in the N-$1s$ 
orbital.\cite{dal2014pseudopotentials} In order eliminate spurious interactions 
between a core-hole and its periodic images, we have considered a distance of 
$\sim$7\,\AA\, between the  core-holes.

The electronic structure calculations  and structural/thermal stability 
simulations  were performed using the VASP code. We have considered an energy 
cutoff of 500\,eV for the plane wave basis set, and the Brillouin zone was 
sampled using  a  15$\times$15$\times$1 k-point mesh.\cite{mp} The structural 
stability was verified through the calculation of elastic constants and the 
phonon dispersion using PHONOPY code.\cite{togoScrMat2015} The 
thermal stability was verified by ab initio molecular dynamics simulations 
(AIMD) at 300K, with a time step of 1 fs using Nos\'e heat bath 
scheme.\cite{JCP81-511-1984}

The calculation of the electronic transmission probability ($T$) 
 was performed based on the non-equilibrium Green's functions (NEGF) 
formalism using the DFT Hamiltonian as implemented in the Siesta and 
TranSiesta\cite{siesta,transiesta} codes. The KS orbitals were expanded in  a 
linear combination of numerical pseudo-atomic orbitals, using split-valence 
double-zeta basis set including polarization functions.\cite{dzp, EShiftSiesta}  
The BZ samplings were performed using two different set of k-point meshes, 
$1\times 10\times 200$  $1\times 20\times 500$ according with the  
electronic transport  directions.

The total  transmission probability of electrons with energy $E$ and 
bias voltage $V$, $T(E,V)$, from the left electrode to reach the right 
electrode passing through the scattering region is given by, 
$$
T\left(E \right) = Tr \left[\Gamma_{\mathrm R}\left(E,V\right) G^{\mathrm 
R}\left(E,V\right) \Gamma_{\mathrm L}\left(E,V\right) G^{\mathrm 
A}\left(E,V\right) \right], 
$$ 
where $\Gamma_{L,(R)} 
\left(E,V\right)$ is the coupling with the left and right electrodes and 
$G^{R,(A)}$ is the retarded (advanced) Green function matrix of the scattering 
region. The current $I$ is evaluated by using Landauer-B\"uttiker 
formula,\cite{buttikerPRB1985,landauer1988generalizd}
$$
I\left(V\right)=\frac{2e}{h}\int 
T\left(E,V\right)\left[f\left(E-\mu_L\right)-f\left(E-\mu_R\right) 
\right]dE,
$$
where $f\left(\epsilon\right)$ is the Fermi-Dirac distribution for energy 
$\epsilon$ and $\mu_{L(R)}$ is the electrochemical potential of left 
(right) electrode. We have considered the zero-bias approximation, 
$T(E,V)\approx T(E,0)$, for the calculation of the  electronic current 
 calculated at the limit of low bias voltage ($\le$\,0.1\,V).

\section{Results and Discussions}

\subsection{Pristine $A_2B$ electrenes}

\begin{figure}
    \includegraphics[width=8cm]{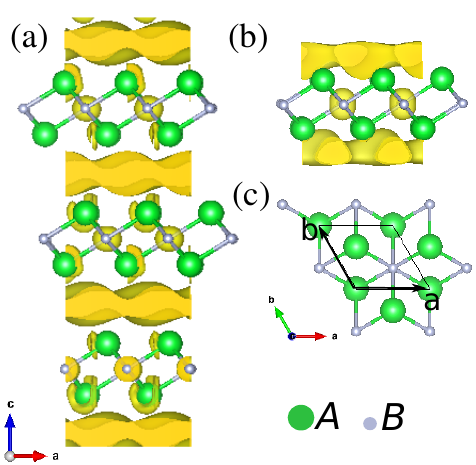}
    \caption{\label{models1}Structural model of \ab\, electride bulk (a), and  
single layer electrene,  side view (b) and top view (c).  The isosurfaces (of 
0.003\,$e$/\AA$^3$) show the localization of the anionic electrons within the 
energy interval of  $\pm 0.5$\,eV with respect to the Fermi level.}
\end{figure}

The $A_2B$ electrides with   $A$\,=\,Ca, Sr, Ba, Y, and  $B$\,=\,N, P, As, C 
share the same structure of \can, Fig.\,\ref{models1}(a). Our results of 
equilibrium geometries of $A_2B$ electrides and 2D single layer electrenes, 
summarized in Table\,I, are in good agreement with previous experimental and 
theoretical findings, viz.: \baas.\cite{mingJACS2016}, \bap,\cite{mingJACS2016}, 
\srp,\cite{mingJACS2016}, \yc,\cite{zhangChemMat2014,houJPhysChemC2016} 
\can,\cite{gregoryJMatChem2000,wangMaterials2018,qiuJPhysChemC2019,wuJMMM2020} 
\srn,\cite{breseJSolStateChem1990,wuJMMM2020} and 
\ban.\cite{reckewegZeits2005,wuJMMM2020} The structural properties of these 
electrides are anisotropic, characterized by a strong intralayer interactions 
due  to the  $A$--$B$ ionic chemical bonds, and comparatively weaker interlayer 
interaction between the \ab\, sheets.  The latter is ruled by  a superposition 
of (i) Coulombic attractive forces between the positively charged \ab\, layer 
and the anionic electrons, and (ii) repulsive interaction between the positively 
charged \ab\, layers.\cite{zhaoJACS2014, druffelJACS2016,druffelJMatChemC2017} 
In order to provide a quantitative picture of the interlayer binding strength, 
we calculate the interlayer binding energy ($E^b$) defined 
as,\cite{jungNanoLett2018} 
%%%%
$$
E^b=\frac{1}{S}\left( E[A_2B]_{\rm ML}-E[A_2B]_{\rm Bulk} \right),
$$ 
%%%%
where $E[A_2B]_{\rm ML}$  and $E[A_2B]_{\rm Bulk}$ are the total energies of 
single layer electrene, and  $A_2B$  electride, respectively, and $S$ is the 
surface area normal to the \ab\, stacking.  Our  results of  binding energies 
for \can, \srn, \ban, and \yc\, (Table\,I) are in good agreement with those 
presented  in the current literature.\cite{dalePCCP2017,liuJPhysChemC2020}

\begin{table}
 \centering
  \caption{\label{key} Details of the equilibrium geometry of $A_2B$ electrides 
 and single layer electrenes, lattice constant $a$ and {\it A--B} equilibrium 
bond length (in \AA), and the interlayer binding energy, $E^{b}$ 
(in J/m$^2$) without/with the inclusion of vdW interactions.}
\begin{ruledtabular}
  \begin{tabular}{lccccc}
 & \multicolumn{2}{c}{{bulk}}&               &\multicolumn{2}{c}{{monolayer}}\\
$A_2B$ &   $a$    &{\it A--B}&$E^b$&{\it a}  &{\it A--B}\\
\cline{1-1}      \cline{2-3}   \cline{4-4}       \cline{5-6} 
 \baas\,    & 4.64&    3.21      & 0.41/0.48     & 4.65    &   3.22 \\
\bap\,     & 4.65&    3.18      & 0.44/0.51     & 4.64    &   3.17  \\
 \srp\,     & 4.45&    3.01      & 0.59/0.64     & 4.43    &   3.00 \\
 \yc\,      & 3.61&    2.47      & 1.60/1.73     & 3.50    &   2.45 \\
 \can\,     & 3.60&    2.42      & 0.97/1.02     & 3.61    &   2.43 \\
 \srn\,     & 3.84&    2.60      & 0.78/0.83     & 3.85    &   2.61 \\
  \ban\,     & 4.02&    2.76      & 0.58/0.64     & 4.00    &   2.75 \\
 \end{tabular}
\end{ruledtabular}
\end{table}

\begin{figure}
    \includegraphics[width= \columnwidth]{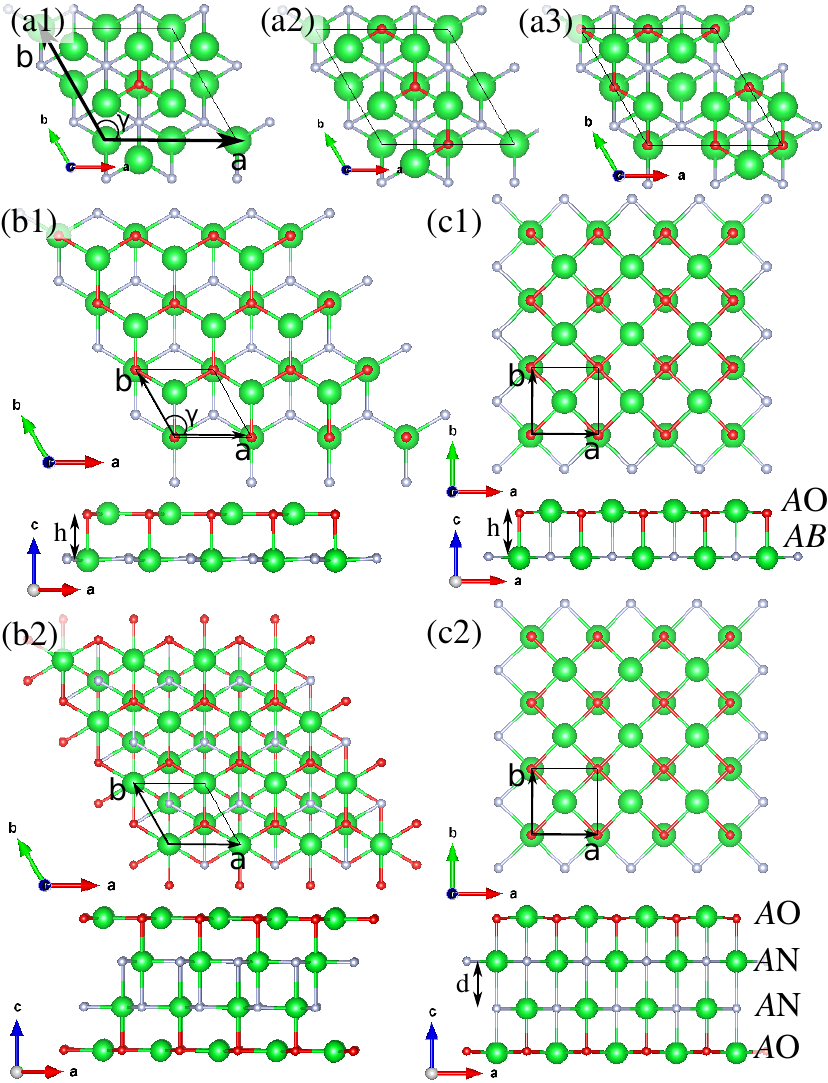}
    \caption{\label{models2}Structural models of the hexagonal electrene adsorbed 
    by O adatoms, with oxygen  coverage ($n$) of 0.25 (a1), 0.50 (a2), and 0.75 
(a3).    Hexagonal (b1) and tetragonal (c1) one-sided full oxidized single 
layer electrene; two-sided full oxidized (b2) hexagonal and (c2) tetragonal 
bilayer electrene. Oxygen atoms are indicated by red spheres.}
\end{figure}

Given the large \ab--\ab\, interlayer distance ($>3$\,\AA), it is worth to 
examine the contribution of the van der Waals (vdW) interactions in the binding 
energies. As shown in Table\,I, the  calculations of $E^b$ without/with the 
inclusion of the vdW interactions reveal a slight increase  of $E^b$, for 
instance between 5 an 10\% for the nitrides, $A_2$N. Thus, we can infer that the 
Coulombic attractive forces bring the major contribution to the interlayer 
interactions.

Our results of  $E^b$ indicate that these \ab\, electrenes can be classified as 
``potentially exfoliable'' based on the criteria presented, by Mounet {\it et 
al.}, in a recent high-throughput computational investigation applied to two 
dimensional material.\cite{mounetNatNanotech2018} As shown in 
Figs.\,\ref{models1}(b) and (c), at the equilibrium geometry, the single layer 
electrene  exhibits the same \ab\, atomic structure of its bulk (electride) 
parent, with  the  anionic electrons lying on the electrene surface.

\subsection{Oxidation}

Here we will address the energetic stability, structural characterization,  and 
the electronic properties of the oxidized \ab\, electrenes. Firstly, we have 
considered the one-sided full oxidized single layer electrene [\oab], and in 
the sequence the two-sided full oxidized bilayer electrene [\oababo]. In 
Figs.\,\ref{models2}(b1) and (b2) we present the structural models of \oab, and \oababo.

\begin{figure}
    \includegraphics[width= \columnwidth]{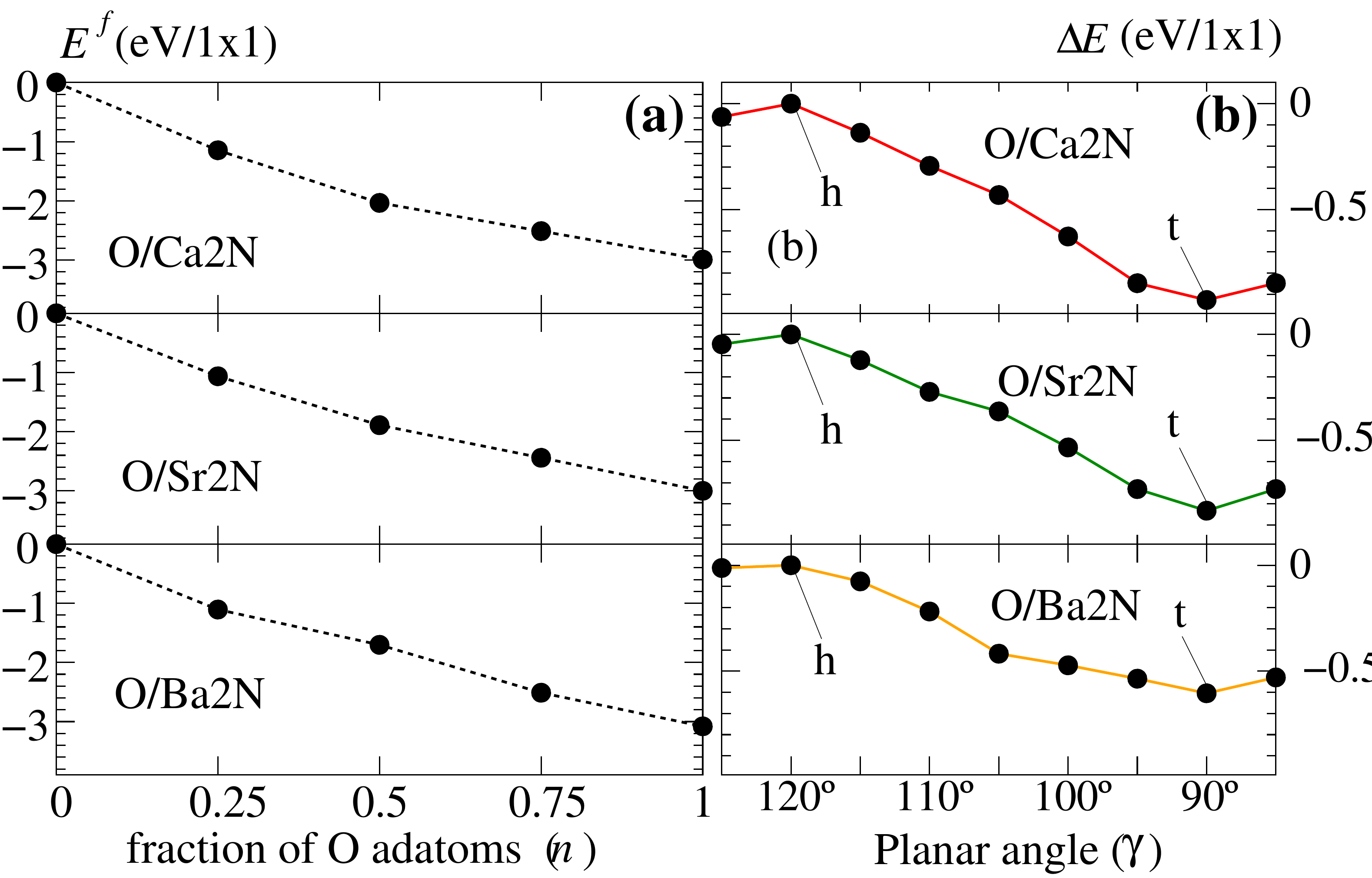}
    \caption{\label{energy-neb}(a) Formation energy of the oxidized \oab\, 
electrene,  and (b) the  total energy difference as a function of the in-plane 
angle ($\gamma$)  formed by the lattice vectors {\sf a} and {\sf b} 
[Figs.\,\ref{models2}(b1)] of the fully oxidized electrenes, 
\oab, with $\gamma$\,=\,120$^\circ$ for the hexagonal (h) phase and 90$^\circ$ 
for the tetragonal (t) phase.}
\end{figure}

\subsubsection{Energetic Stability}

\begin{figure*}
    \includegraphics[width=2\columnwidth]{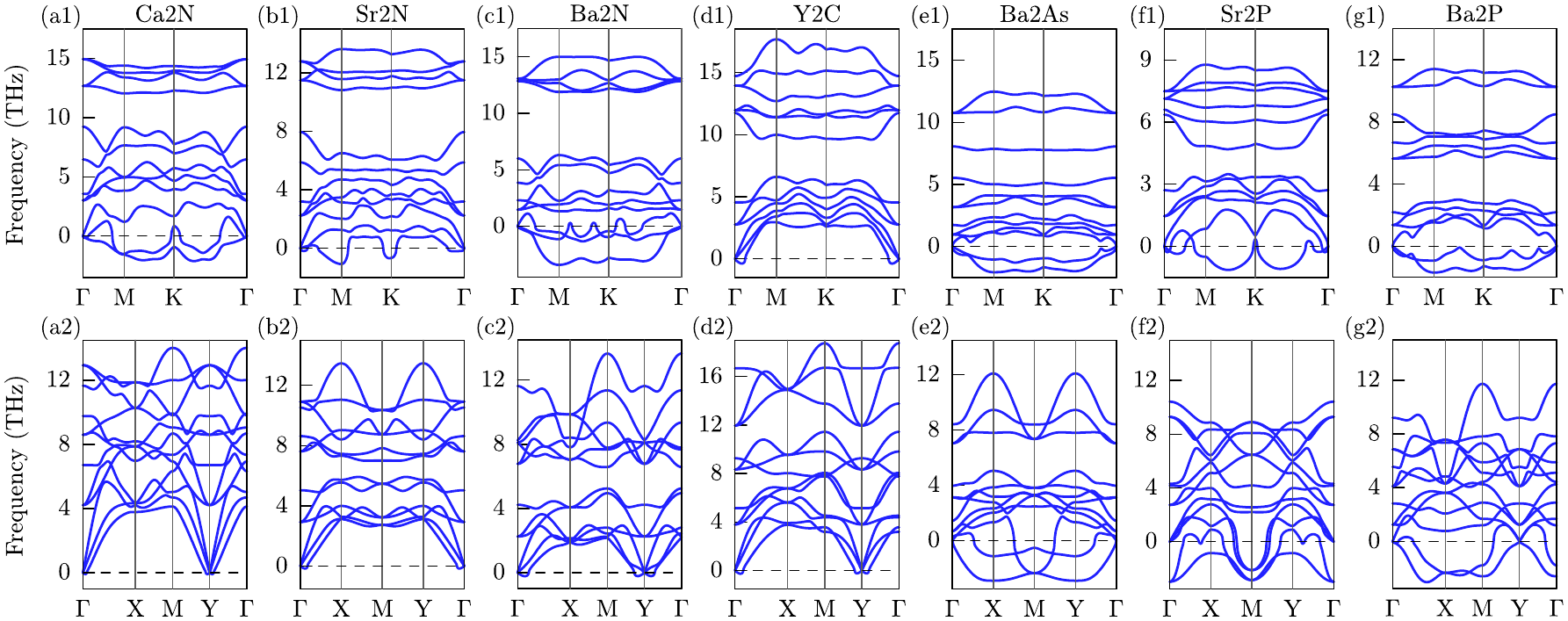}
    \caption{\label{phonons-1ml} Phonon spectra of the hexagonal (a1)-(g1), and 
tetragonal (a2)-(g2)  \oab\, electrenes.}
\end{figure*}

The energetic stability of the oxidized electrenes was inferred through  the 
calculation of the formation energy ($E^f$), 
$$ 
E^f =  E[{\rm O}_n/A_2B] - E[A_2B] - 2n\,E[{\rm O_2}].
$$ 
E$[{\rm O}_n/A_2B]$ and $E[A_2B]$ are the total energies of the oxidized 
and pristine \ab\, electrenes, where $n$ is the fraction of O adatoms per 
2\,$\times$\,2 surface unit cell, as shown 
Figs.\,\ref{models2}(a1)-(a3).  The total energy calculations were 
performed by considering the full relaxation of the atomic positions, and the 
lattice vectors {\sf a} and {\sf b} in Fig.\,\ref{models2}; $E[{\rm O_2}]$  is 
the total energy of an isolated O$_2$ molecule (triplet state).

We found an energetic preference for oxygen adatoms on the hollow site aligned 
with the (cation) $A$ atom at the opposite side of the $A_2B$ monolayer, as 
shown in Figs.\,\ref{models2}(a1)-(a3) and \ref{models2}(b1) for $n$\,=\,0.25, 
0.50, 0.75, and 1, respectively.\cite{souzaJPhysChemC2020,liJPhysChemC2020} In Table\,IV\, (Appendix), we present a summary of our total energy results for oxygen adsorption on the other surface sites, namely aligned with the (cation) $B$ atom, and on-top of the $A$ atom of the same surface side. Our formation energy results indicate  that the oxidation processes are exothermic, with  
$E^f$ (in absolute values) proportional to the oxidation rate. In 
Fig.\,\ref{energy-neb}(a) we present $E^f$ as a function of the oxygen coverage 
for \oab\, with  $B$\,=\,N, \oaN. It is worth pointing out  that, although the 
anionic electrons of \ab\, are neutralized for coverage of 50\%  
(O$_{0.5}$/\ab),  the  incorporation of oxygen adatoms for $n$\,$>$\,50\% is 
energetically favorable. For instance, O$_{0.5}$/\can\,  plus an excess of O$_2$ 
molecules is less stable than the full oxidized O/\can\, system by 
0.96\,eV/O-atom, O$_{0.5}$/\can\,$+$\,O$_2$\,$\xrightarrow{-0.96}$\,O/\can. 
Similar results were obtained for the other \oab\, systems. The formation 
energies, and the equilibrium lattice constant of \oab\, are summarized in 
Table\,II. 

However, although the negative values of $E^f$, phonon spectra calculations 
[Fig.\,\ref{phonons-1ml}] revealed  imaginary frequencies for all hexagonal 
\oab\, structures [hereinafter referred to as \oabh], except O/(\yc)$^\text{h}$ 
[Fig.\,\ref{phonons-1ml}(d1)], thus indicating that the other \oabh\, systems 
are dynamically unstable. We have also examined  structural stability of the 
hexagonal phases through molecular dynamics (MD) simulations, where found that 
all \oabh\, systems, with an exception for  O/(\yc)$^\text{h}$, are structurally 
unstable [Fig.\,\ref{md-1ml}, Appendix].

In the sequence, we performed a search for dynamically and structurally 
stable O/\ab\, by changing the in-plane angle ($\gamma$)  formed by the 
lattice vectors {\sf a} and {\sf b} [Figs.\,\ref{models2}(b1)]. Such a 
structural search was guided by the existence of stoichiometrically equivalent   
$A$O and $A$N tetragonal (t) bulk cubic parents, namely CaO, SrO, and 
BaO,\cite{dadsetaniSolStatSci2009, nejatipourPhysScrip2015, nguyenPhysLettA2021, 
MatProj} and CaN, SrN, and BaN.\cite{MatProj} Our findings, for $B$\,=\,N, show 
that indeed the full oxidized hexagonal geometry [Fig.\,\ref{models2}(b1)], 
$\gamma$\,=\,120$^\circ$,   corresponds to a metastable configuration 
characterized by a barrierless (h\,$\rightarrow$\,t) structural transition  
[Fig.\,\ref{energy-neb}(b)] to a tetragonal ($\gamma$\,=\,90$^\circ$) 
$A$O/$A$N layered phase, \oaNt\, [Fig.\,\ref{models2}(c1)].

\begin{table}
 \centering
  \caption{\label{key} Formation  energy ($E^f$ in eV/1\,$\times$\,1), 
for $n$\,=\,1, of the hexagonal  
and tetragonal oxidized  single layer (\oab), and bilayer [\oababo] electrenes, 
and  the total energy gain upon hexagonal\,$\rightarrow$\,tetragonal structural 
transition ($\Delta E^{\text{h-t}}$ in eV/O-atom). The lattice constant 
($a$) and the vertical distances (d and h in  
Fig.\,\ref{models2}) are in \AA. The lattice constant of the pristine 
hexagonal $A_2B$ electrene are within parentheses.}
\begin{ruledtabular}

  \begin{tabular}{cccccc}
    \multicolumn{1}{c}{{O/$A_2B$}}    & \multicolumn{2}{c}{hexagonal} 
& \multicolumn{1}{c}{{$\longrightarrow$}}               & 
\multicolumn{2}{c}{tetragonal}  \\
$A_2B$  & $E^f$&      {\it a}& $\Delta E^{\text{h-t}}$ & {\it a} & 
h 
  \\
\cline{1-1} \cline{2-3} \cline{4-4} \cline{5-6}  
\can\, & $-2.99$ &     3.85 (3.61)   & $-0.90$  & 3.36     &  2.40 \\
 \srn\, & $-3.00$ &     4.11 (3.85)   & $-0.80$  & 3.59    & 2.54 \\
 \ban\, & $-3.08$ &     4.35 (4.00)   & $-0.57$  & 3.83    & 2.61   \\
  \yc\,  & $-5.46$ &     3.64 (3.50)   & $-0.20$  & 3.40  &  2.58 \\
 \baas\,& $-1.79$ &     4.07 (4.65)   & $-0.51$  & 3.75   & 2.80 \\
  \bap\, & $-2.08$ &     4.42 (4.64)   & $-0.63$  & 4.18   & 2.64  \\
 \srp\, & $-2.21$ &     4.33 (4.43)   & $-0.76$  & 3.92    & 2.50 \\
 \hline
    \multicolumn{1}{c}{{O/$(A_2B)_2$/O}}    & \multicolumn{2}{c}{hexagonal} 
& \multicolumn{1}{c}{{$\longrightarrow$}}               & 
\multicolumn{2}{c}{tetragonal}  \\
 $A_2B$  & $E^f$&      {\it a}& $\Delta E^{\text{h-t}}$ & {\it a} &  
d   \\
\hline
\can\, & $-3.04$  &  3.92 & $-1.68$  & 3.40 & 2.57     \\
\srn\, & $-3.07$  &  4.19 & $-1.27$  & 3.63 & 2.85     \\
\ban\, & $-3.02$  &  4.33 & $-1.10$  & 3.86 & 3.22     \\
  \end{tabular}
\end{ruledtabular}
\end{table}
% 
%  \can\,     & 3.60&          & 0.97/1.02     & 3.61    &    \\
%  \ban\,     & 4.02&          & 0.58/0.64     & 4.00    &    \\
%  \bap\,     & 4.65&          & 0.44/0.51     & 4.64    &    \\
%  \srn\,     & 3.84&          & 0.78/0.83     & 3.85    &    \\
%  \srp\,     & 4.45&          & 0.59/0.64     & 4.43    &    \\
%  \yc\,      & 3.61&          & 1.60/1.73     & 3.50    &    \\
%  \baas\,    & 4.64&          & 0.41/0.48     & 4.65    &    \\

% 
% \begin{figure*}
%     \includegraphics[width=2\columnwidth]{md-paper.png}
%     \caption{\label{md-1ml} MD simulations of   O/$A_2B$ electrenes.}
% \end{figure*}

\begin{figure}
    \includegraphics[width=\columnwidth]{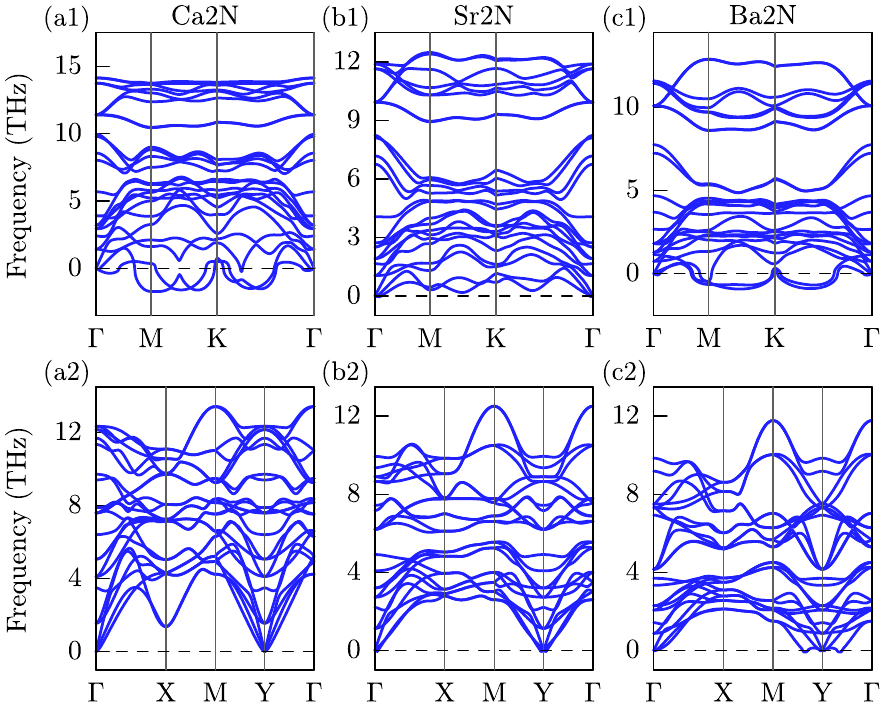}
    \caption{\label{phonons-2ml} Phonon spectra of the hexagonal (a1)-(c1), and 
tetragonal (a2)-(c2)  \oaNaNo\,  electrenes.}
\end{figure}

The energy gain upon such a h\,$\rightarrow$\,t  transition is given 
by  the total energy difference between the two structural phases, 
$
\Delta E^{{\text{t-h}}}=E^{\rm t} - E^{\rm h}.
$
Our results of $\Delta E^{{\text{t-h}}}$, presented in Table\,II, indicate that 
the tetragonal phase, \oabt, is energetically more favorable for all \oab\, 
systems. On the other hand, phonon spectra calculations revealed that the 
\oabt\, systems are  dynamically stable only for $B$=N [\oaNt] and  (YOYC)$^{\rm 
t}$.  As shown in Fig.\,\ref{phonons-1ml}, imaginary frequencies present in 
the hexagonal \oaN\, structures [Figs.\,\ref{phonons-1ml}(a1)-(c1)] were 
suppressed in the tetragonal \ocant,  \osrnt, and \obant\, systems, 
Figs.\,\ref{phonons-1ml}(a2)-(c2). In addition, the structural stability of  
\oaNt\, and  (YOYC)$^{\rm t}$ was confirmed  by  MD simulations. We found that 
the tetragonal structure was preserved at a temperature of 300\,K during the 
15\,ps of MD simulation [Fig.\,\ref{md-1ml} (Appendix)].
\begin{figure}
    \includegraphics[width=\columnwidth]{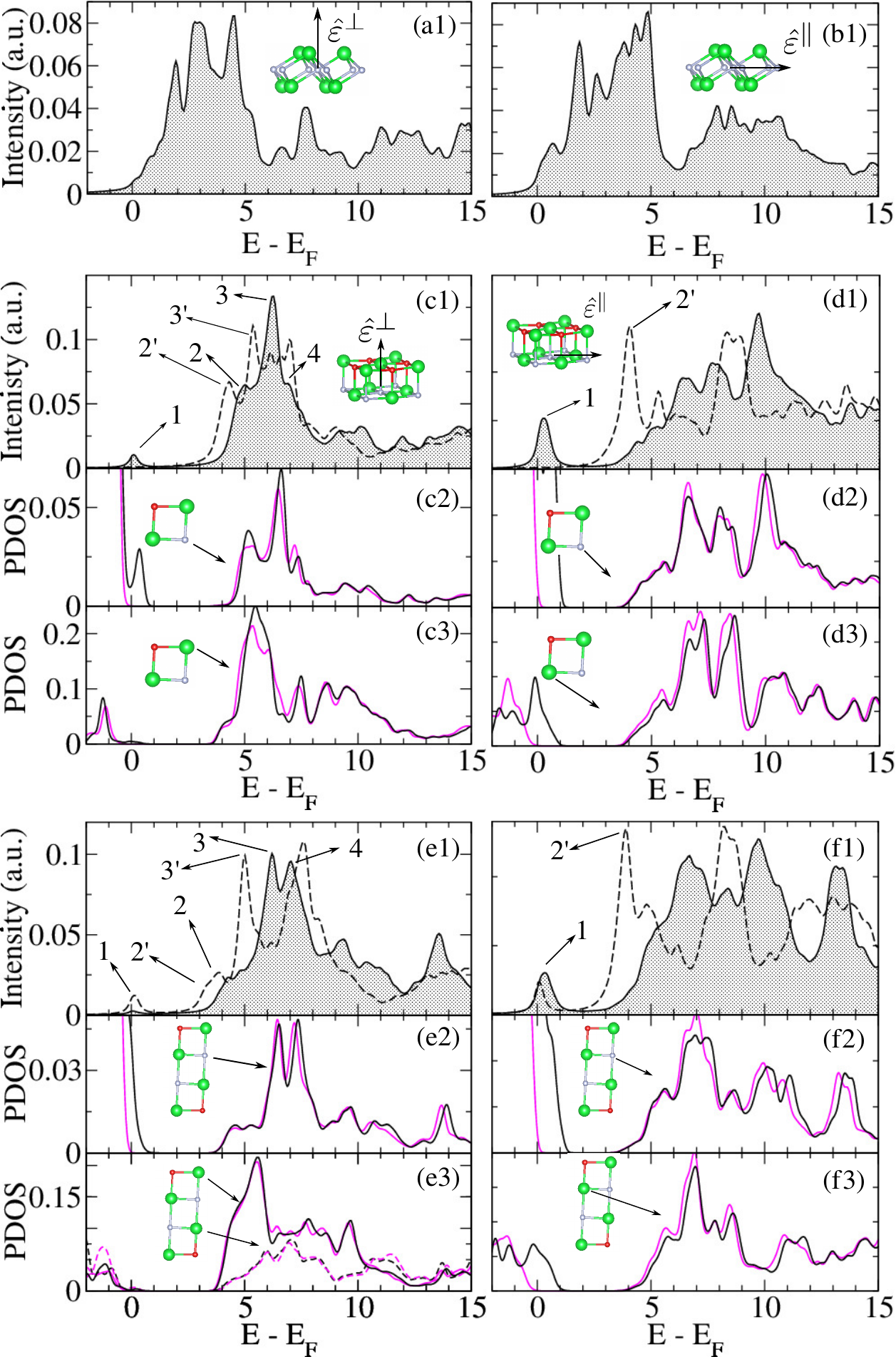}
    \caption{\label{xanes} XANES spectra of pristine single layer \can\,
electrene for the polarization vector (a1) perpendicular, 
$\hat\varepsilon^\perp$, and (b1) parallel,  $\hat\varepsilon^\parallel$, to 
the electrene surface. XANES spectra of \ocan\, for $\hat\varepsilon^\perp$ 
(c1),  $\hat\varepsilon^\parallel$ (d1), and the density 
of states of \ocant\, projected  on the N-$2p_{\rm z}$ (c2), Ca-$4p_{\rm z}$ 
(c3), N-$2p_{\rm x,y}$ (d2), and Ca-$4p_{\rm x,y}$ (d3) orbitals. XANES spectra 
of \ocancano, for $\hat\varepsilon^\perp$ 
(e1),  $\hat\varepsilon^\parallel$ (f1), and the density 
of states of \ocancanot\, projected on the N-$2p_{\rm z}$ (e2), 
Ca-$4p_{\rm z}$ (e3), N-$2p_{\rm x,y}$ 
(f2), and Ca-$4p_{\rm x,y}$ (f3) orbitals. The XANES spectra of the tetragonal 
(hexagonal) phase are indicated by solid (dashed) lines. Spin-down and spin-up 
channels are indicated by purple and black solid lines.}
\end{figure}

We next have examined the surface oxidation of bilayer  electrenes [\abab] with 
$B$\,=\,N, \oaNaNo\, [Fig.\,\ref{models2}(b2)]. Similarly to what we have found 
in the single layer systems,  the \oaNaNo\, structures are (i) energetically 
stable ($E^f<0$ in Table\,II), and also (ii)  present  exothermic 
h\,$\rightarrow$\,t structural transitions, with $\Delta E^{\text{t-h}}$ almost 
twice compared with those of  their \oab\, counterparts. Further phonon spectra 
calculations, and molecular dynamics simulations of \oaNaNo, 
Figs.\,\ref{phonons-2ml} and \ref{md-2ml} (Appendix), respectively, show that 
the tetragonal (layered) systems, depicted in Fig.\,\ref{models2}(c2), are 
dynamically and structurally stable. We found no imaginary frequencies in the 
tetragonal phases, Figs.\,\ref{phonons-2ml}(a2)-(c2), and  the MD simulations 
reveal that the atomic structures of the \oaNaNot\, systems have been preserved 
[Figs.\,\ref{md-2ml}(a2)-(c2), Appendix], whereas the ones of the hexagonal 
phase are no longer maintained after 15\,ps of simulation, 
Figs.\,\ref{md-2ml}(a1)-(c1) (Appendix).

At the equilibrium geometry, the  tetragonal phase of the oxidized electrenes is 
characterized by  a layered structure indicated as $A$O and $A$N in 
Fig.\,\ref{models2}(c2), where the inner $A$N bilayer structure, with interlayer 
bond distance d (indicated in Table\,II), is shielded by oxidized $A$O sheets, 
\oaNaNot. As expected, such geometries somewhat mimic the ones of their 
(stoichiometrically equivalent and energetically stable)  $A$O and $A$N bulk 
cubic parents, namely CaO, SrO, and BaO,\cite{dadsetaniSolStatSci2009, 
nejatipourPhysScrip2015, nguyenPhysLettA2021, MatProj} and CaN, SrN, and 
BaN,\cite{MatProj} thus providing further support to the  energetic and 
structural stability of the tetragonal \oaNaNot.

\subsubsection{Structural Characterization}

In order to present a more complete structural picture of the oxidized 
systems, in connection with their electronic properties, and also provide a 
theoretical support for future experimental studies
we have simulated the nitrogen K-edge X-ray absorption spectra of 
pristine ($A_2$N), and  the oxidized \oaN\, and  \oaNaNo\, systems.  Here we 
will  present our results for $A$\,=\,Ca, namely \can, \ocan\, and 
\ocancano; the other systems, with $A$\,=\, Sr and Ba, present quite similar 
spectra and interpretations. 

Let us start with  the single layer pristine \can\,  electrene.  In 
Figs.\,\ref{xanes}(a1) and (b1) we present the absorption spectra  for 
polarization vector perpendicular ($\hat\varepsilon^\perp$)  and   parallel 
($\hat\varepsilon^\parallel$) to the surface, respectively. Based on the 
analysis of orbital projected density of states (DOS, not shown), we found  that 
the edge and near-edge absorption features are mostly dictated by the electronic 
transition from the N-$1s$ core electron to the unoccupied N-$2p_{\rm z}$ and 
N-$2p_{\rm x,y}$ orbitals, for $\hat\varepsilon^\perp$ and  
$\hat\varepsilon^\parallel$, respectively. Due to the electronic confinement 
along the normal direction with respect to the electrene surface,  the 
broadening of the absorption lines from the Fermi energy ($E_{\rm F}$) up to  
$\sim$\,$E_{\rm F}+6$\,eV for $\hat\varepsilon^\perp$ is slightly  smaller 
compared with that for $\hat\varepsilon^\parallel$. Meanwhile, in  the oxidized 
systems   the  energy broadenings for  $\hat\varepsilon_{\parallel}$ 
[Figs.\,\ref{xanes}(d1) and (f1)] are significantly larger compared with those  
for $\hat\varepsilon_{\perp}$, Figs.\,\ref{xanes}(c1)-(e1),  indicating a 
reduction (increase) of the electronic confinement along  the parallel  
(perpendicular) direction with respect to the surface plane;  which is,  in its 
essence, a consequence of the formation of planar Ca--N and Ca--O layered 
structures upon oxidation [Fig.\,\ref{models2} and insets of 
Figs.\,\ref{xanes}(c1) and (d1)]. Further identification of the oxidized 
structures can be done by comparing the energy position of the absorption edges 
(AEs). For instance, comparing the AEs of the tetragonal phases, indicated by 
solid lines in Figs.\,\ref{xanes}(a1) and (c1), we find that  the former  lies 
near the Fermi level, while in the latter the AE starts at about $E_{\rm 
F}+4$\,eV, thus, indicating an increase of the N-$1s$ binding energy (BE) in the 
oxidized systems. Indeed, based on the L$\rm\ddot{o}$wdin charge population 
analysis, we found that the total charge of the nitrogen atoms in the oxidized 
\ocan\, [\ocancano] electrenes reduces by 0.34 [0.42]\,$e$/N-atom when compared 
with the one  of pristine \can. Thus,  we can infer  that the increase of the 
N-$1s$ BE is  due to the reduction of the electronic screening at the N nucleus 
in the oxidized \can.

Next, we examine the nitrogen K-edge XANES spectra of the oxidized tetragonal 
systems in light of the projected density of states. The  projections  on the 
N-$2p_{\rm z}$ and Ca-$4p_{\rm z}$ orbitals, Figs.\,\ref{xanes}(c2) and (c3), 
indicate  that  the absorption features 2, 3,  and 4 in Fig.\,\ref{xanes}(c1) 
are  ruled by the electronic transitions  to the lowest unoccupied N-$2p_{\rm 
z}$ (major contribution) hybridized with the nearest neighbor Ca-$4p_{\rm z}$  
orbitals (minor contribution). It is worth noting that the features 2 and 3 (for 
$\hat\epsilon^\perp$) of the tetragonal phase are also present in the absorption 
spectra  of the hexagonal phase [2' and 3' (dashed lines) in 
Fig.\,\ref{xanes}(c1)]. However they are shifted by $\sim$1\,eV toward lower 
energies when compared with their counterparts 2  and 3, thus we can infer  that 
the BE of the N-$1s$ core electrons in the tetragonal phase is larger compared 
with the one  of the hexagonal phase. As discussed above, such an increase of 
the BE is supported by the  reduction of the total charge  of the nitrogen atoms 
(by 0.01\,$e$/N-atom) in the tetragonal \ocan\, in comparison with that of 
hexagonal one. Similarly for \ocancano, as shown Figs.\,\ref{xanes}(e1)-(e3), 
(i) the XANES spectra ($\hat\epsilon^\perp$) of the tetragonal phase is ruled by 
the unoccupied N-$2p_{\rm z}$ states (major contribution) hybridized with the 
$4p_{\rm z}$ orbitals (minor contribution) of the Ca atoms embedded in the CaO 
sheets; and  (ii) the edge features of the tetragonal and hexagonal phases 
indicate that the BE of N-$1s$ core electrons of the former is larger by about 
1\,eV\, compared with that of the latter, in agreement with the lower total 
charge (by 0.04\,$e$/N-atom) of the N atoms in the tetragonal phase.

In Figs.\,\ref{xanes}(d) and (f) we present the XANES spectra for a polarization 
vector parallel to the \ocan\, and \ocancano\, layers, 
$\hat\varepsilon^\parallel$, and the  DOS projected on the N-$2p_{\rm x,y}$ and 
Ca-$4p_{\rm x,y}$ orbitals of the tetragonal phase. Compared  with the 
absorption spectra with the polarization vector normal to the surface, 
$\hat\varepsilon^\perp$, we found that the  pre-edge absorption (feature 1),  
attributed to the  hybridizations of the partially occupied spin-down N-$2p_{\rm 
x,y}$ and Ca-$4p_{\rm x,y}$ orbitals, becomes more intense for 
$\hat\varepsilon^\parallel$. In fact, such a pre-edge absorption spectrum can be 
considered as a signature of the formation of half-metallic channels along the 
CaN layers, which will be discussed below. In the sequence, it is noticeable the 
well defined absorption spectrum 2', present in the hexagonal phase, has been 
practically suppressed in the tetragonal structure, thus suggesting that the 
structural differences between the tetragonal and hexagonal phases are better 
captured by looking at the in-plane edge absorption features.

% 
%                               angle   
%                            O-A-O  A-B-A
% O/Ba2As: BaO=2.70/a=3.75,   159    111 
% O/Ba2P:  BaO=2.97/a=4.18,   170    170
% O/Sr2P:  SrO=2.77/a=3.92,   176    156
% O/Y2C:   YO =2.41/a=3.40,   174    168
% O/Ba2N:  BaO=2.72/a=3.83,   167    173
% O/Ca2N:  CaO=2.38/a=3.36,   178    180
% O/Sr2N:  SrO=2.54/a=3.59,   173    177
%
% Id    Formula	Spacegroup FormEn  EAboveHull BandGap	Volume	Nsites	Density
%mp-1342 BaO	Fm3m	 -2.831	      0	      2.256	44.255	 2	5.753
%AB = 2.81 angs; a=3.97
%mp-2605 CaO	Fm3m	 -3.314	      0	      3.692	28.332	 2      3.287
%AB = 2.42 angs, a=3.42
%mp-2472 SrO	Fm3m	 -3.085       0       3.449	35.277	 2	4.878
%AB = 2.60 angs, a=3.68
%mp-1187855 YO	P63/mmc  -3.103    0.214      0.000	67.536	 4	5.159
%%%%%%%%%%%%%%%%%%%%%%%
%mp-1057758 BaN Fm3m	-0.133	   0.517      0.000	46.91	 2	5.357
%AB = 2.86       a=4.05   FM, 1\mu_B
%mp-1058549 CaN Fm3m	-0.475	   0.398      0.000	31.547	 2	2.847
%AB = 2.51       a=3.55   FM, 1\mu_B
%mp-1058586 SrN Fm3m	-0.281	   0.434      0.000	39.119	 2	4.314
%AB =  2.69      a=3.81   FM, 1\mu_B
%mp-998893 YC	Fm3m	 0.043	    0.36      0.000	33.041	 2	5.072
%AB = 2.55       a=3.60

\subsubsection{Electronic and Magnetic Properties}

The electronic band structures of  \ab\, electrenes, with $A$\,=\,Ca, Sr, 
Ba, and $B$\,=\,N are characterized by  parabolic metallic bands, giving rise 
to nearly free electron (NFE) states localized on the electrene's surface and 
between the  stacked layers. On the other hand, upon the formation of $A$O 
oxidized layers these NFE states become unoccupied, and we observe  the 
emergence of magnetic moments in the $A$N layers. Here, we will examine  the  
magnetic  and electronic properties of \oaNt\,  and \oaNaNot.

\begin{figure}
    \includegraphics[width=\columnwidth]{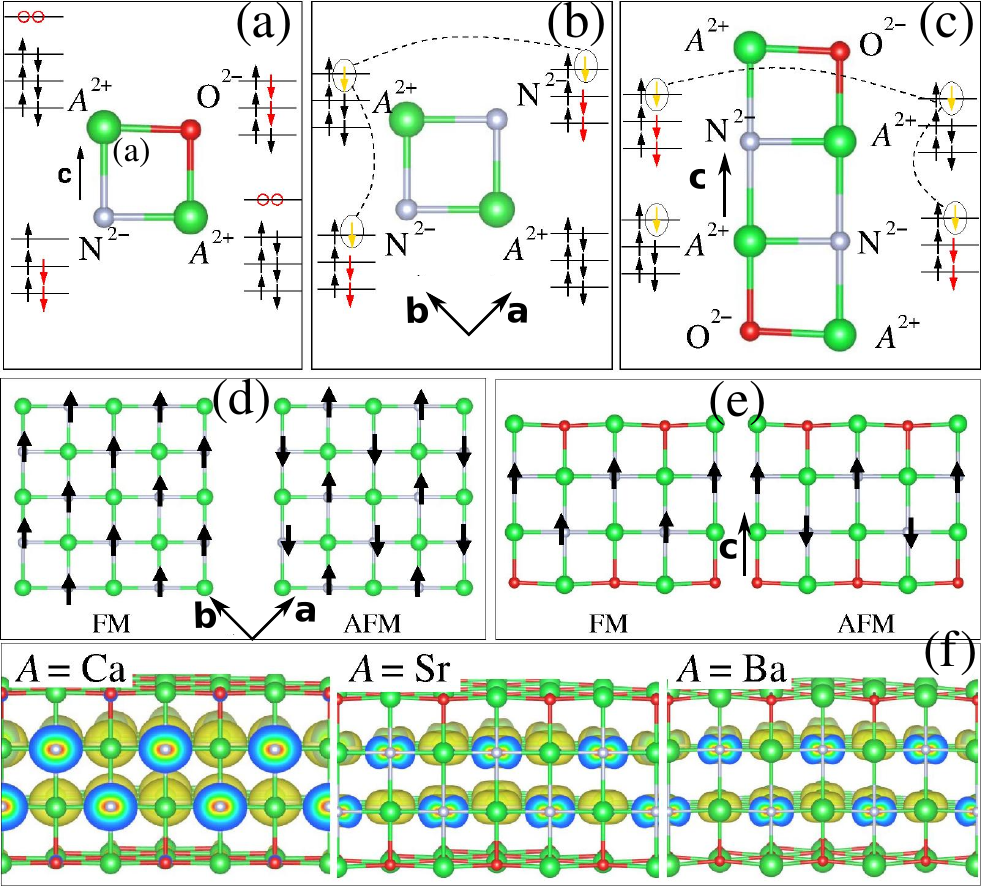}
    \caption{\label{mag-scheme} Schematic orbital occupation of 
\oaNt\, (a) $A$O and NO layers along the stacking direction ({\sf c}), and (b) 
$A$N sheet perpendicular to the stacking direction, {\sf a}\,$\times$\,{\sf b} 
plane. (c) Orbital occupation of \oaNaNot\, along the stacking direction.
(d) Intralayer and (e) interlayer FM/AFM spin-polarizations. (f) Spin-density 
distribution of the (intralayer and interlayer) FM \oaNaNot\, systems. 
Isosurface of 0.004\,$e$/\AA$^2$.}
\end{figure}

\begin{figure}
    \includegraphics[width=\columnwidth]{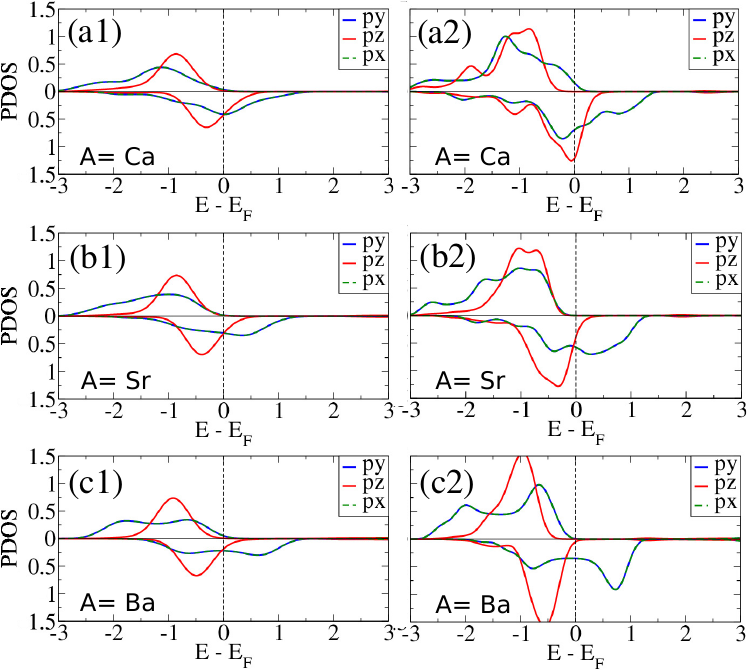}
    \caption{\label{pdos} Electronic density of states (DOS) projected on the 
N-$2p$ orbitals of \oaNt\, (a1)-(c1), and \oaNaNot\, (a2)-(c2).}
\end{figure}

% 
%       vasp-gga       vasp-D2     qe-D2     ox-hex
% Ba2N   3.78 eV        3.972       8.41
% Ba2P   3.45 eV        4.028       3.72   -2.11770
% Sr2N   4.12 eV        4.198       3.89   -2.92565
% Sr2P   3.27 eV        3.613       3.28   -2.33422
% Y2C    6.06 eV        6.085       5.69   -5.49352
% Ba2As  3.71 eV        4.638       4.33   -2.62151
% 
%  \can\,     &     &          &               & 3.57    &  2.41  \\
%  \ban\,     &     &          &               & 3.99    &  2.76  \\
%  \bap\,     &     &          &               & 3.92    &  2.89  \\
%  \srn\,     &     &          &               & 3.76    &  2.57  \\
%  \srp\,     &     &          &               & 4.39    &  2.99  \\
%  \yc\,      &     &          &               & 3.50    &  2.43  \\
%  \baas\,    &     &          &               & 3.97    &  2.98  \\

Based on the nominal oxidation states and the electronegativities of the 
involved atoms, we can infer the emergence of  a magnetic moment in the oxidized 
systems.\cite{wuJMMM2020} There is a net charge transfer of 2 electrons from 
each (less electronegative) $A$ atoms to the (more electronegative) O and N 
atoms, resulting in $A^{2+}$, O$^{2-}$ and N$^{2-}$ oxidation states, 
Fig.\,\ref{mag-scheme}(a).  The ground state configuration  is characterized by  
an O$A$ (oxidized) layer with closed $p$ shells parallel to a $A$N layer with 
the N-2$p$  orbitals partially occupied,  Fig.\,\ref{mag-scheme}(b). Similarly, 
in the bilayer system, \oaNaNot\, [Fig.\,\ref{mag-scheme}(c)],  we find  $A$N 
layers with partially occupied N-2$p$ orbitals sandwiched by O$A$ (edge) layers 
with closed $p$ shells. According to the Hund's rule, each N atom will carry a 
net magnetic moment of 1\,$\mu_{\rm B}$. Indeed, within the GGA-PBE approach, we 
found a  net  magnetic moment of about 0.8\,$\mu_{\rm B}$  mostly localized on 
the nitrogen atoms.\cite{vasp-check} The projected electronic density of states 
(PDOS)  on the N-$2p$ orbitals of \oaNt\, and \oaNaNot, Figs.\,\ref{pdos}(a) and 
(b), reveal that the partial occupation of planar N-$2p_{\rm x,y}$ orbitals 
brings the major contribution to the polarization of the N atoms. 

Total energy comparisons between  the magnetic and non-magnetic phases, $\Delta E^{\rm mag}=E^{\text{mag}}-E^{\text{non-mag}}$, support the energetic preference for the spin-polarized systems, $\Delta E^{\rm mag}$\,$<$\,0 in Table\,III. The strength of the magnetic interactions between the nitrogen atoms was examined by comparing the total energies of the ferromagnetic (FM) and antiferromagnetic (AFM) phases for the intralayer ($\Delta E^{\text{FM-AFM}}_{\rm intra}$), and interlayer ($\Delta E^{\text{FM-AFM}}_{\rm inter}$) magnetic couplings,   as shown in Figs.\,\ref{mag-scheme}(d) and (e), respectively. The intralayer coupling   takes place   between the N atoms  in the same $A$N layer  [Fig.\,\ref{mag-scheme}(d)], while the interlayer coupling   is due to the  interactions between the N atoms lying in different  $A$N layers  of  \oaNaNot, Fig.\,\ref{mag-scheme}(e). Our results, summarized in Table\,III, reveal the both systems, \oaNt\, and \oaNaNot, present an energetic preference for the intralayer and interlayer FM coupling between the N atoms. It is noticeable  that (i) \ocancanot\, presents the largest interlayer FM interaction, $\Delta E^{\text{FM-AFM}}_{\rm inter}$\,=\,$-36.5$\,meV/N-atom, when compared with the other \oaNaNot\, systems, leading to (ii) a strengthening of the  intralayer FM  coupling, namely $\Delta E^{\text{FM-AFM}}_{\rm intra}$\,=\,$-8.5$\,$\rightarrow$\,$-22.8$\,meV/N-atom. In contrast, (iii) we found (relatively)  lower values of $\Delta E^{\text{FM-AFM}}_{\rm inter}$ for \osrnsrnot\, and \obanbanot\, which can be due to the larger interlayer distance (d) as indicated in Fig.\,\ref{models2}(c2) and Table\,II, and the more localized feature of the spin-polarized states normal to the stacking direction ({\sf c}).  In Fig.\,\ref{mag-scheme}(f) we present the spin-density distribution of the intralayer and interlayer FM \oaNaNot, with $A$=Ca, Sr, and Ba where we confirm the localization of the net magnetic moment on the nitrogen atoms. It is noticeable that the projection of the DOS on the N-$2p$ orbitals [Fig.\,\ref{pdos}] support the larger interlayer interaction ruled  by the N-$2p_{\rm z}$ orbitals in \ocancanot\, [Fig.\,\ref{pdos}(a2)] compared   with those of the other oxidized bilayer electrenes, Figs.\,\ref{pdos}(b2) and (c2).

The energetic preference for the FM phase can be attributed to super-exchange 
interactions between the N$^{2-}$ anions mediated by $A^{2+}$ cations. In this 
sense, the FM coupling will be favored due to the electron  delocalization along 
the N$^{2-}$--$A^{2+}$--N$^{2-}$ bonds,  thus lowering the kinetic energy of the 
system, as schematically shown in Figs.\,\ref{mag-scheme}(b) and (c)  for the 
intralayer and interlayer couplings, respectively.  Further support to the FM 
coupling between the N$^{2-}$ anions, mediated by super-exchange interactions, 
can be found in the Goodenough-Kanamori 
rule,\cite{goodenoughPRB1955,kanamoriJAP1960} since  the 
N$^{2-}$--$A^{2+}$--N$^{2-}$ bonds are characterized by bond angles of 
90$^\circ$.

\begin{table}
 \centering
  \caption{\label{key} Total energy differences (in meV/N-atom) between 
non-magnetic and magnetic phases, $\Delta E^{\rm 
mag}=E^{\text{mag}}-E^{\text{non-mag}}$, and between the FM and AFM phases for 
intralayer  ($\Delta E^{\text{FM-AFM}}_{\rm intra}$), and interlayer ($\Delta 
E^{\text{FM-AFM}}_{\rm inter}$) interactions, schematically shown in 
Figs.\,\ref{mag-scheme}(d) and (e), respectively.}
\begin{ruledtabular}
  \begin{tabular}{lrrr}
    %\multicolumn{1}{c}{$A_2B$}    & \multicolumn{1}{c}{$\Delta E^{\rm mag}$ } 
    %& \multicolumn{1}{c}{$\Delta E^{\text{FM-AF}}_{\rm intra}$}  & 
%\multicolumn{1}{c}{$\Delta E^{\text{FM-AF}}_{\rm intra}$} \\
         &  $\Delta E^{\rm mag}$ &  $\Delta E^{\text{FM-AFM}}_{\rm intra}$ & 
$\Delta E^{\text{FM-AFM}}_{\rm inter}$    \\  
           \hline                    
\ocant\,     & $-67$  \hspace{3mm}  & $-8.5$ \hspace{3mm} &  -- \hspace{5mm} \\
\osrnt\,     & $-135$ \hspace{3mm} & $-18.7$ \hspace{3mm} & --  \hspace{5mm} \\
\obant\,     & $-92$  \hspace{3mm} & $-1.9$  \hspace{3mm} &  -- \hspace{5mm}  \\
\hline
\ocancanot\,    & $-83$ \hspace{3mm} & $-22.8$ \hspace{3mm} & $-36.5$ 
\hspace{3mm}  \\
\osrnsrnot\,    & $-127$ \hspace{3mm} & $-12.9$ \hspace{3mm} &  $-1.0$ 
\hspace{3mm} \\
\obanbanot\,    & $-93$ \hspace{3mm}  & $-4.4$ \hspace{3mm} &  $-0.5$ 
\hspace{3mm}  \\
  \end{tabular}
\end{ruledtabular}
\end{table}

\begin{figure}
    \includegraphics[width=\columnwidth]{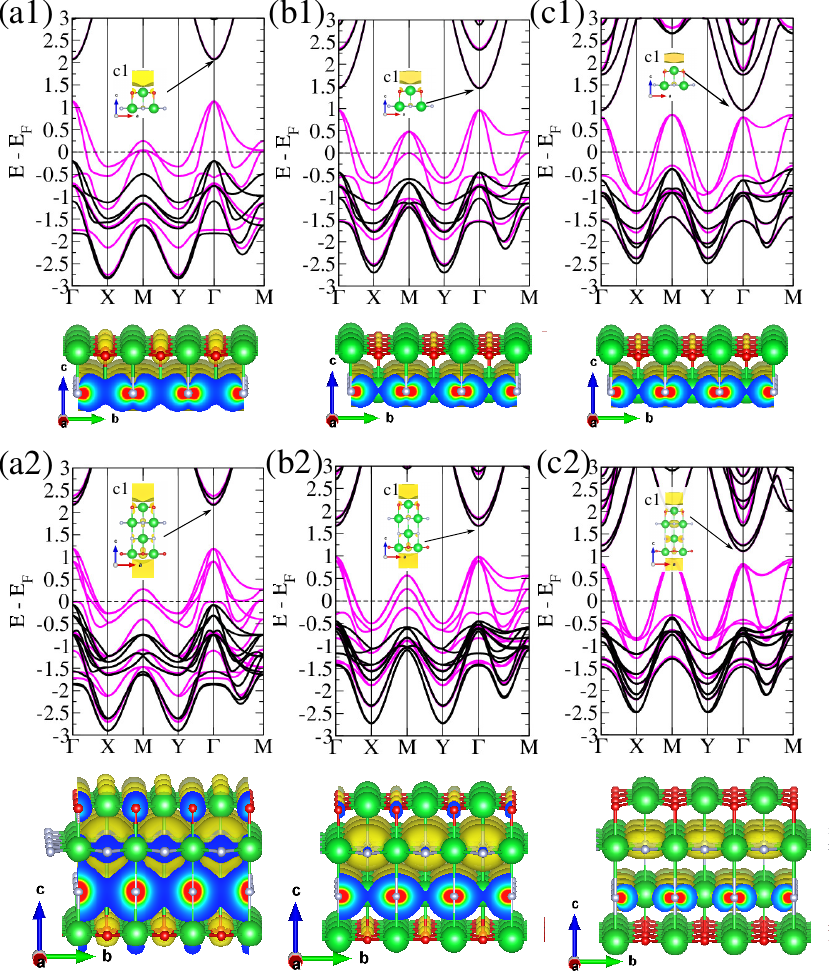}
    \caption{\label{bands} Electronic band structure of and the 
electronic distribution near the Fermi level ($E_{\rm F}\pm 0.1$\,eV) 
of the oxidized \oabt\, monolayer with  $A$\,=\, Ca (a1),  Sr (b1), and Ba (c1), and 
the oxidized \oababot bilayer with $A$\,=\, Ca (a2), Sr (b2), and Ba (c2). Electronic distribution of the NFE state at the $\Gamma$-point (inset). Isosurfaces of 0.002\,$e$/\AA$^2$. Solid purple and black lines indicate the spin-down and spin-up channels.}
\end{figure}

The   electronic band structures of \oaNt\, and \oaNaNot, Fig.\,\ref{bands}, indicate they are half-metals. The metallic channels are characterized by spin-down (purple-lines), whereas the spin-up energy bands (black-lines) are semiconductor with the valence band maximum (VBM) lying at about 0.2\,eV below the Fermi level ($E^{\rm VBM}$\,$\approx$\,$E_{\rm F}-0.2$\,eV) for $A$\,=\,Ca [Fig.\,\ref{bands}(a1)-(a2)] while for $A$\,=\,Sr and Ba we find $E^{\rm VBM}$\,$\approx$\,$E_{\rm F}-0.5$\,eV, Figs.\,\ref{bands}(b1)-(b2) and (c1)-(c2). The lowest unoccupied states are spin degenerated,   lying between 1 and 2\,eV above $E_{\rm F}$, and characterized by NFE parabolic bands localized on the oxidized surface (O$A$) [insets of Fig.\,\ref{bands}]. Further real space projections of the electronic states near the Fermi level, $E_{\rm F}\pm 0.1$\,eV, reveal that  the half-metallic bands are mostly ruled by in-plane N-2$p$ orbitals localized in the $A$N layers of \oaNt,  Figs.\,\ref{bands}(a1)-(c1) . Similarly,  in the bilayer systems the half-metallic bands spread out through the $A$N layers; however, it is worth noting that in \oaNaNot, these half-metallic channels are sandwiched by the oxidized $A$O sheets  [Figs.\,\ref{bands}(a2)-(c2)]. These oxidized sheets may act as a shield,  protecting the half-metallic channels against the environment conditions, which is a quite appealing property for development of spintronic devices based on 2D platforms.

\begin{figure}
    \includegraphics[width=\columnwidth]{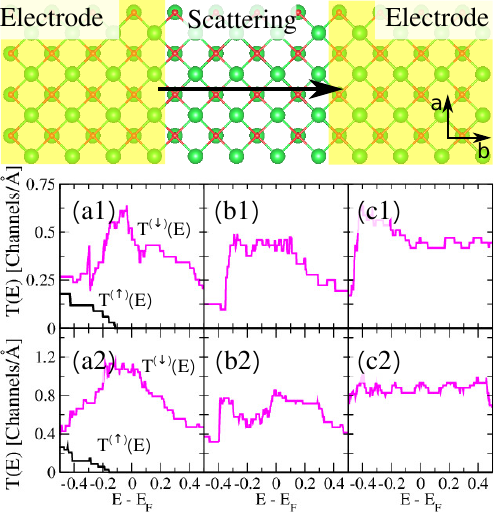}
    \caption{\label{fig-trans} (top) Structural model of the simulation setup 
used for the electronic transport calculation along the {\sf b} direction. 
Transmission probability and electronic current of the \oabt\, oxidized 
electrene with  $A$\,=\,Ca (a1),  Sr (b1), and Ba (c1), and \oababot\, with  
$A$\,=\,Ca (a2),  Sr (b2), and Ba (c2).}
\end{figure}

In order to provide a quantitative picture of the emergence of spin-polarized  
electronic current in the oxidized electrenes, we calculate the electronic 
transmission probability, $T(E)$, and the electronic  current $I(V)$  of the  
\oaNt, and \oaNaNot\, systems. In Fig.\,\ref{fig-trans}(top) we present the 
simulation setup used for the electronic transport calculations along the {\sf 
b} direction, namely two (left and right) electrodes composed by semi-infinite 
oxidized electrenes, both connected to a (central) scattering region. Our 
results of $T(E)$, summarized in Figs.\,\ref{fig-trans}(a1)-(a3) and (b1)-(b3), 
reveal that  near the Fermi level ($|E_\text{F}|\leq 0.1\,\text{eV}$) the 
transmission probability is mediated by the spin-down channels, in consonance 
with the electronic band structure results [Fig.\,\ref{bands}]. The transmission 
channels lie on the inner $A$N layers, mostly ruled by the in-plane 
hybridizations between the spin-down N-$2p$ orbitals 
(N-$2p_\text{x,y}^{(\downarrow)}$) with the nearest neighbor $A$ atoms. For 
$A$\,=\,Ca  we found larger values of $T(E)$ [Figs.\,\ref{fig-trans}(a1) and 
(a2)], and net electronic current [Figs.\,\ref{fig-curr}(a1) and (a2) 
Appendix] when compared with the other  \oaNt\, and \oaNaNot\, systems. In 
contrast, although the more localized character of the in-plane  
N-$2p_\text{x,y}^{(\downarrow)}$ orbitals for $A$\,=\,Ba [Figs.\,\ref{bands}(c1) 
and (c2)] compared with the ones obtained for $A$\,=\,Sr, Figs.\,\ref{bands}(b1) 
and (b2), we found nearly the same values of transmission probability, 
Figs.\,\ref{fig-trans}(b1)-(b2) and (c1)-(c2)], and  electronic current for low 
bias voltage [Figs.\,\ref{fig-curr}(b1)-(c1) and (c1)-(c2) Appendix]. We 
have also calculated the transmission probability and the electronic current 
along the bisector direction between {\sf a} and {\sf b}, where we found 
practically the same values of $T(E)$ and $I(V)$ near the Fermi level. Thus, 
suggesting that the electronic transport through the inner $A$N layers, at low 
bias limit,  does not present significant directional anisotropy as predicted in 
the other  2D systems.\cite{padilhaPCCP2016}

\section{Summary and Conclusions}

By means of first-principles DFT calculations, we have performed a theoretical study of the full oxidized 2D single layer, \oab,  and bilayer, \oababo, electrenes, with $A$\,=\,Ba, Ca, Sr, Y, and $B$\,=\, As, N, P, C. We found that \oab\, and \oababo\, systems  with $A$\,=\,Ca, Sr, Ba , and $B$\,=\,N become stable  upon an hexagonal\,$\rightarrow$\,tetragonal structural transition, resulting in layered tetragonal systems, \oaNt\, and \oaNaNot. Further characterizations,  through simulations of XANES spectroscopy, allowed us to identify key aspects of the absorption spectra and their correlation with the structural and electronic properties the oxidized systems. We found the emergence of a ferromagnetic phase in the oxidized tetragonal structures,  with the net magnetic moment mostly ruled by the planar N-$2p_{\rm x,y}$ orbitals. Meanwhile, electronic structure calculations reveal the formation of half-metallic bands spreading out through  the $A$N layers, with nearly negligible contribution from the oxidized $A$O sheets. The emergence of spin-polarized transmission channels was confirmed through the electronic transport calculations based on the Landauer-B\"uttiker formalism. These results reveal   that the oxidized \oaNt\, and \oaNaNot\, systems are  quite interesting platforms for spin-polarized transport on 2D systems, characterized spin-polarized metallic channels shielded by oxide layers. For instance,  \oaNaNot\, can be viewed as a core-shell 2D material with half-metallic channels lying on the $(A\text{N})_2$ layers (core) protected against the environment conditions by the oxidized $A$O sheets (shell).

\begin{acknowledgments}

The authors acknowledge financial support from the Brazilian agencies CNPq, 
CAPES, FAPEMIG, and INCT-Nanomateriais de Carbono, and the CENAPAD-SP and Laborat{\'o}rio Nacional de Computa{\c{c}}{\~a}o Cient{\'i}fica (LNCC-SCAFMat2) for computer time.

\end{acknowledgments}
 \section{Appendix}

In Table\,\ref{sites} we present the total energy differences between the \oab\, electrenes with the oxygen atom adsorbed on the (i)\,hollow site aligned with the cation $A$ atom at the opposite surface side of the \ab\, monolayer, and (ii)\,on the  hollow site aligned with the $B$ atom, and (iii)\,on-top of the $A$ atom of the same surface side, namely   $\Delta E({\rm ii})=E({\rm i)}-E({\rm ii})$, and  $\Delta E({\rm iii})=E({\rm i)}-E({\rm iii})$, respectively. Our results of $\Delta E$, viz.: $\Delta E({\rm ii})<0$ and $\Delta E({\rm iii})<0$, confirm the energetic preference of (i).

 \begin{table}
   \caption{\label{sites} Total energy differences (in eV/O-atom) between the (energetically most stable) O adatom on the hollow site  aligned with the $A$ atom at the opposite surface side of the \ab\, monolayer [$E({\rm i})$], and the O adatom on the  hollow site aligned with the $B$ atom [$E({\rm ii})$], and the O adatom on the on-top site aligned with  $A$ atom of the same surface side, $\Delta E({\rm ii})=E({\rm i)}-E({\rm ii})$, and  $\Delta E({\rm iii})=E({\rm i)}-E({\rm iii})$, respectively.} 
   \begin{ruledtabular}
   \begin{tabular}{lcc}
\oab\,  &  $\Delta E({\rm ii})$ &   $\Delta E({\rm iii})$  \\
 \hline
 O/\can  &   $-0.12$          &  $-2.82$ \\    
 O/\srn  &    $-0.19$          & $-2.77$    \\
 O/\ban   &  $-0.34$           & $-2.60$       \\
 O/\yc   &   $-0.33$         &   $-2.91$    \\
 O/\baas  &  $-0.12$         &   $-2.82$ \\
 O/\srp    & $-0.15$        &    $-3.37$    \\
 O/\bap    & $-0.19$      &      $-2.75$   \\
   \end{tabular}
 \end{ruledtabular}
   \end{table}

 \begin{figure*}
    \includegraphics[width=2\columnwidth]{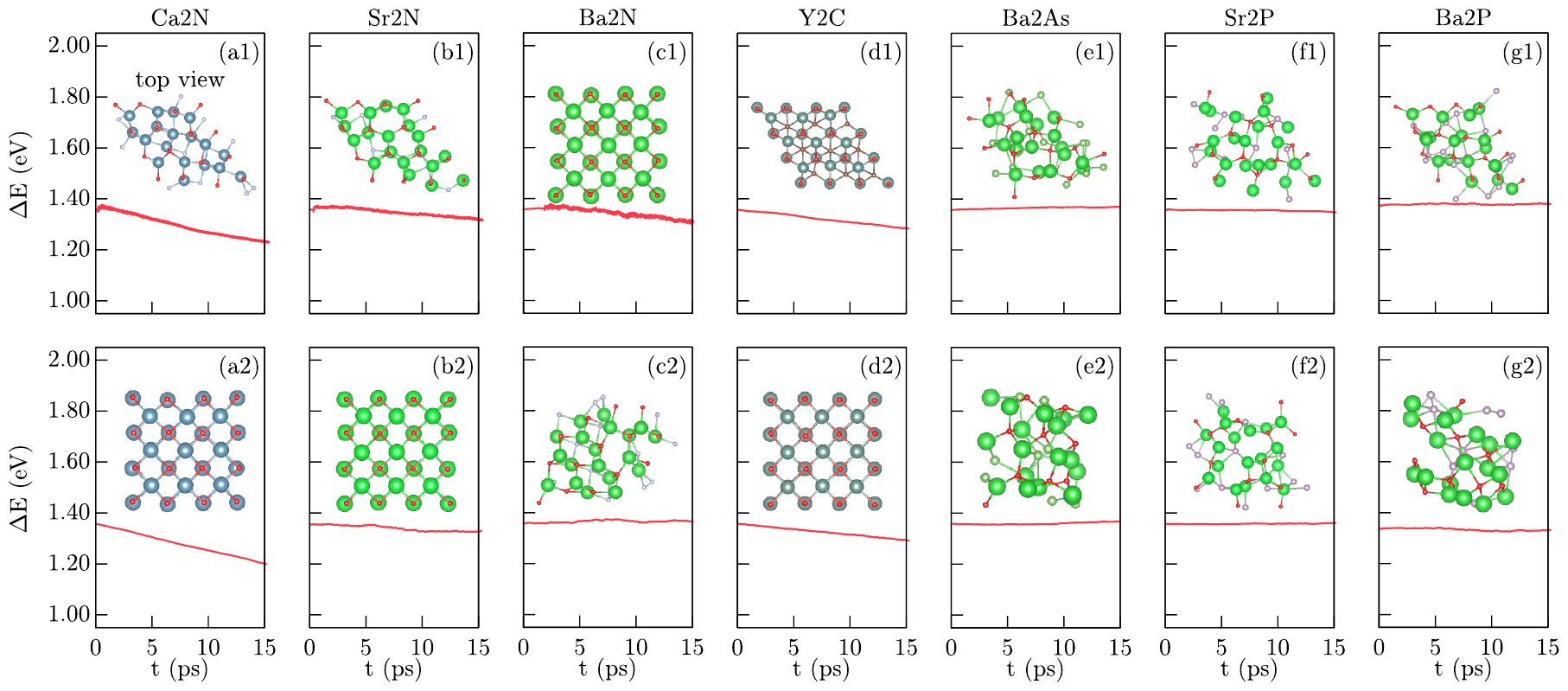}
    \caption{\label{md-1ml} Total energy fluctuation,  of 
hexagonal (a) and  tetragonal (b) \oab\, oxidized electrenes, as a function of 
the time step (1\,fs).  Insets, strutural model after 15\,ps of molecular 
dynamics simulation at 300\,K.}
\end{figure*}
 
\begin{figure}
    \includegraphics[width=\columnwidth]{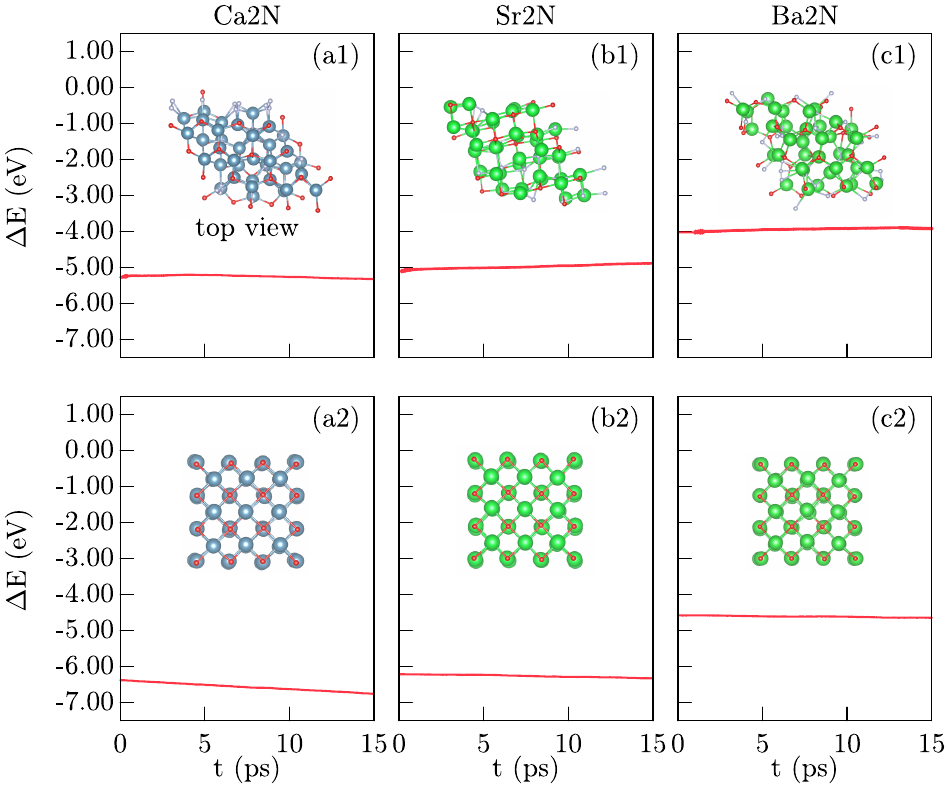}
    \caption{\label{md-2ml}  Total energy fluctuation,  of 
hexagonal (a) and  tetragonal (b) \oaNaNo\, oxidized electrenes, as a function 
of the time step (1\,fs).  Insets, strutural model after 15\,ps of molecular 
dynamics simulation at 300\,K.}
\end{figure}

In Figs.\,\ref{md-1ml} and \ref{md-2ml} we present our results of MD simulation 
o oxidized single layer electrene, \oab, with $A$= A = Ca, Sr, 
Ba, Y, and $B$ = N, P, As, C, and bilayer (\oaNaNo) electrenes, with $A$=Ca, Sr 
and Ba. We have considered a total simulation time of 15\,ps, and time steps of 
1\,fs. Inset we present the strutural model after 15\,ps of simulation at 
300\,K.
 
% 
% \begin{figure}
%     \includegraphics[width=7cm]{bands-nomag.jpg}
%     \caption{\label{bands-nomag} Electronic band structure of \baas, \bap, 
% \srp\, and and \yc.}
% \end{figure}

\begin{figure}
    \includegraphics[width=\columnwidth]{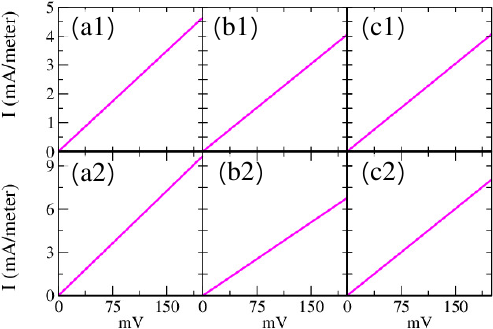}
    \caption{\label{fig-curr}  Electronic current of  \oabt\ with $A$\,=\,Ca (a1), Sr (a2), and Ba (a3), and \oababot\, with $A$\,=\,Ca (a2), Sr (b2), and Ba (c2).}
\end{figure}

In Fig.\,\ref{fig-curr} we present the net electronic current of the oxidized monolayer \oabt, and bilayer \oababot\, systems by using Landauer-B\"uttiker formula as described in Section\,II.

\bibliography{RHMiwa.bib}

\begin{thebibliography}{10}
\bibitem{vasp-check}
We have verified the accuracy of our total energy results by using the VASP
  code, by using $15\times 15\times 1$ special k-points\cite{mp} to the
  Brillouin zone sampling, and energy cutoff of 600\,eV for the plane wave
  basis set.
\end{thebibliography}

\end{document}